\documentclass[12pt]{article}
\usepackage{graphicx, url}
\usepackage{palatino}
\usepackage{color}
\usepackage{amsmath, amssymb,bbm}
\usepackage{epsfig}
\usepackage{rotating}
\usepackage{dcolumn}
\usepackage{bm}
\usepackage{caption}
\usepackage{subcaption}
\usepackage[T1]{fontenc}
\usepackage[table]{xcolor}
\newcommand{\comment}[1]{}

\def \bea{\begin{eqnarray}}
\def \eea{\end{eqnarray}}

\def \ccbar{c\overline{c}}

\begin{document}   
\baselineskip 18pt
\title{{Predictions for $h_c$ and $h_b$ production at the LHC}}
\author{
   Sudhansu~S.~Biswal$^1$\footnote{E-mail: sudhansu.biswal@gmail.com}, 
   ~Sushree~S.~Mishra$^1$\footnote{E-mail: sushreesimran.mishra97@gmail.com},\\ 
   ~Monalisa Mohanty$^1$\footnote{E-mail: monalimohanty97@gmail.com}
     ~and  K.~Sridhar$^2$\footnote{E-mail: sridhar.k@apu.edu.in} \\ [0.2cm]
    {\it \small 1. Department of Physics, Ravenshaw University,} \\ [-0.2cm]
    {\it \small Cuttack 753003, India}\\ [-0.2cm]
    {\it \small 2. School of Arts and Sciences, Azim Premji University,} \\ [-0.2cm]
    {\it \small Sarjapura, Bangalore 562125, India}\\
}
\date{}
\maketitle
\begin{abstract}
\noindent  
The production cross sections of $h_c$ and $h_b$, the ${}^1P_1$ quarkonia can be predicted in 
Non-Relativistic Quantum Chromodynamics (NRQCD) 
using heavy quark symmetry. Our study includes predictions for 
both the integrated cross-section and the transverse
momentum ($p_T$) distribution of $h_c$ ($h_b$) production at the Large Hadron Collider (LHC), 
using the decay process 
$h_c (h_b) \rightarrow \eta_c (\eta_b) + \gamma$, $\eta_c (\eta_b) \rightarrow p \overline{p}$. 
We demonstrate substantial discrepancies in integrated cross-section and 
the $p_T$ distribution of ${}^1P_1$ quarkonia production at the LHC
using Colour-Singlet Model (CSM), NRQCD and modified NRQCD. 
Measuring these resonances at the LHC could discriminate
between these models, thereby offering further insights into the dynamics of
quarkonium production. In addition, we compare the recent LHCb data
for the integrated cross-section
of $h_c$ production at $\sqrt{s}$ = 13 TeV in the kinematic range 
5.0 $<$ $p_T$ $<$ 20.0 GeV and 2.0 $<$ $y$ $<$ 4.0 
with the theoretical predictions 
using NRQCD and modified NRQCD. Modified NRQCD
gives an agreement with the recent LHCb experimental data.
\end{abstract}
\maketitle

\maketitle

\noindent
The discovery of the first quarkonium state, $J/\psi$ in 1974 has significantly expanded 
our understanding of quarkonium properties and the theoretical concepts 
and methods based on Quantum Chromodynamics (QCD).
Potential models~\cite{Barnes:2005pb} explain all charmonium states 
below the $D\overline{D}$ threshold, as observed experimentally.  
The final charmonium state among these charmonium states below the 
$D \overline{D}$ threshold has been experimentally verified is the 
P-wave spin-singlet state $h_c$ (${}^1P_1$). In 1992, the E760 Collaboration at 
Fermilab \cite{Armstrong:1992ae} first observed the $h_c$ (${}^1P_1$) 
state through $p \bar p$ annihilation.
In recent years, some experimental measurements of P-wave quarkonia, 
including the  $h_c$ and $h_b$ ($1^{+-}$ ) 
(${}^1P_1$ charmonium and bottomonium states) have been conducted.
The corresponding branching ratios \cite{br_hc},
the masses of these quarkonia \cite{CLEO:2005vqq, CLEO:2005ghy, CLEO:2008ero, CLEO:2011aa, Oswald:2013tna} 
and the cross sections for $h_c$ ($h_b$) production through $e^+ e^-$ annihilation at
the CLEO-c \cite{CLEO:2011aa} and B-factories \cite{Oswald:2013tna} have been measured experimentally.
In contrast, only leading-order (LO) results have been provided for  
$h_c$ and $h_b$ production. 
Calculations of $h_c$ hadroproduction at the 
Tevatron \cite{Sridhar:1996vd} and the LHC 
\cite{Sridhar:2008sc, Wang:2014vsa, Qiao:2009zg} predicted a substantial yield.
The photoproduction of $h_c$ was examined in \cite{Fleming:1998md} using a color-octet (CO) 
long-distance matrix element (LDME) derived from the decay 
$B \rightarrow \chi_{cJ} + X$. 
The results suggested a substantial cross section at DESY HERA.
$h_c$ ($h_b$) production via $e^+ e^-$ annihilation \cite{Wang:2012tz, Azizi:2017izn} and by 
the B factory \cite{Bodwin:1992qr, Beneke:1998ks, Jia:2012qx} has been studied. 

The lack of research on $h_c$ ($h_b$) suggests that the 
significance of this meson has been neglected.
First of all, the hadroproduction rate of $h_c$
serves as an excellent test of 
Non-Relativistic Quantum Chromodynamics (NRQCD) ~\cite{bbl}. 
This is due to the fact that NRQCD predictions are substantially 
higher than those based on the Colour-Singlet Model (CSM)~\cite{br, Berger:1980ni}, 
providing an adequate comparison with experimental observations. Additionally, 
the cross sections for $h_c$ 
depend on only one nonperturbative parameter, 
unlike $J/\psi$, where the precise determination of three color-octet long distance
matrix elements introduces ambiguity.
According to the NRQCD scaling rule, the CO LDME for $h_c$ 
must have the same magnitude as $\chi_{c1}$. 
This rule provides an opportunity to test the corresponding 
velocity scaling rule as well as explore $h_c$ 
within the NRQCD framework in spite of the lack of experimental data.

NRQCD has been more successful in explaining the systematics 
of quarkonium production at the Fermilab Tevatron~\cite{cdf,Braaten:2000cm}, 
compared to the then existing CSM, which was used to analyze 
the production of quarkonia, where the $Q \bar Q$ state produced 
in the short-distance process was assumed to be a colour-singlet.
NRQCD predicts transverse polarization at high $p_T$, 
but experiments fail to see any evidence for the polarization 
in $J/\psi$~\cite{jpsi_pol} or $\Upsilon$~\cite{CDF, CMS1, LHCb2} measurements.
Therefore, independent tests of 
NRQCD~\cite{Sridhar:1996vd,Sridhar:2008sc,Braaten:2000cm,tests,Mathews:1998nk,bs1,test1} 
are consequently important and the prediction of polarisation of the produced
quarkonium state is an important test.

Another important test of NRQCD comes from 
the study of quarkonia using heavy-quark symmetry.  
In particular, the non-perturbative parameters 
required for $\eta_c$, $\eta_b$, 
$h_c$ and $h_b$ production can
be obtained, using heavy quark symmetry, 
from the parameters of $J/\psi$, $\Upsilon$, $\chi_c$ and 
$\chi_b$ production respectively. This
approach has been used to predict the 
$\eta_c$, $\eta_b$, 
$h_c$ and $h_b$ production 
cross-sections at the Tevatron~\cite{Sridhar:1996vd,Mathews:1998nk} 
and at the LHC~\cite{Sridhar:2008sc,bs1,Biswal:2023xnk}.
The NRQCD predictions for $\eta_c$ production has shown 
a significant disagreement when compared with 
experimental data from LHCb~\cite{LHCb:2019zaj,LHCb:2014oii}, 
resulting a notable failure.
As there was a major conflict shown between NRQCD predictions 
and observed experimental data,
NRQCD required some modification to address 
quarkonium production fruitfully.

The cross-section for the production of a quarkonium state 
$H$ is given as:
\bea
  \sigma(H)\;=\;\sum_{n=\{\alpha,S,L,J\}} {F_n\over {M}^{d_n-4}}
       \langle{\cal O}^H_n({}^{2S+1}L_J)\rangle, 
\label{factorizn}
\eea
where $F_n$'s are the short-distance coefficients, 
which correspond to the production of $Q \bar{Q}$ in 
the angular momentum and colour state denoted by $n$, 
determined through perturbative QCD calculations. 
The non-perturbative matrix elements,  
${\cal O}_n$ are operaters of naive dimension $d_n$, 
which describe the long-distance physics are not calculable 
and have to be obtained by fitting to available data. 

In a recently proposed modification of 
NRQCD~\cite{Biswal:2023xnk,bms,bms2,bms3}, 
the colour-octet $Q \bar Q$ state can radiate several 
soft $\textit{perturbative}$ gluons -- each emission 
taking away little energy but carrying away units of angular momentum. 
In the multiple emissions that the colour-octet state 
can make before it makes the final NRQCD transition 
to a quarkonium state, the angular momentum
and spin assignments of the $Q \bar Q$ state changes constantly. 
The LHC data on $\eta_c$ production
strongly disagreed with the predictions of 
NRQCD but was in very good agreement
with those of modified NRQCD. 
Moreover the $\eta_b$ production has been predicted in 
both modified NRQCD and NRQCD. In contrast to the case of $\eta_c$ , 
however, the NRQCD predictions for $\eta_b$ 
and the colour-singlet prediction are very similar 
except maybe at very large $p_T$ and 
the modified NRQCD prediction is very different from both these predictions. 
Motivated by these observations, we would like to study $h_c$ and $h_b$
production in both NRQCD and modified NRQCD.
Measurement of these production at the LHC 
will provide a very good test of both NRQCD and modified NRQCD.

The fock-state expansion of quarkonium states is characterized by $v$, 
the relative velocity between quark anti-quark pair.
At leading order, the $Q \bar Q$ state is 
in a colour-singlet state but at $\mathcal{O}(v)$, 
it can be in a colour-octet state and connected to the physical
$h_c$ state through the emission of non-perturbative gluon. 
In NRQCD, the Fock space expansion of the physical $h_c$, which is a $^1P_1$
($J^{PC}=1^{+-}$) state, can be written as: 
\bea
\left|h_c\right>={\cal O}(1)
	\,\left|\ccbar[^1P_1^{[1]}]\right> +	
	{\cal O}(v^2)\,\left|\ccbar[^1S_0^{[8]}]\,g \right>+\cdots ~.
\label{fockexpn}
\eea

In the above equation, the color singlet ${}^1 P_1^{[1]}$ state is 
associated with the order of relative velocity $\mathcal{O}(1)$, 
while the contribution of the color-octet ${}^1 S_0^{[8]}$ 
state occurs at the order $\mathcal{O}(v^2)$.
The color-octet state ${}^1 S_0^{[8]}$ transforms into the 
physical state $h_c$ (${}^1 P_1^{[1]}$) through the emission of a 
gluon in an E1 transition.


The NRQCD cross-section formula for $h_c$ production can be written down 
explicitly in terms of octet and singlet intermediate states. 
The cross-section for $h_c$ production can be represented as:
\begin{eqnarray}
\sigma_{h_{c}} &=& \hat F_{{}^{1}P_1^{[1]}} 
\times \langle {\cal O}^{h_{c}} ({}^{1}P_1^{[1]}) \rangle +  
                  \hat F_{{}^{1}S_0^{[8]}}  
                \times {\langle {\cal O}^{h_{c}} ({}^{1}S_0^{[8]}) } \rangle 
		+ \cdots .
                \label{nrqcd}
\end{eqnarray}

The coefficients $F_n$'s represent the cross sections for 
producing a $\ccbar$ pair in the
angular momentum and color states, denoted by n.
The above NRQCD formula gets modified to the following 
in the modified NRQCD with perturbative soft gluon emission:
\begin{eqnarray}
\sigma_{h_{c}} &=& \hat F_{{}^{1}P_1^{[1]}}   
\times \langle {\cal O}^{h_{c}} ({}^{1}P_1^{[1]})\rangle \cr 
                &+& \biggl\lbrack  
                  \hat F_{{}^{3}S_1^{[8]}} 
                 + \hat F_{{}^{1}P_1^{[8]}} 
                + \hat F_{{}^{1}S_0^{[8]}} 
		+ \hat F_{{}^{3}P_J^{[8]}}  \biggr\rbrack 
		\times ({{\langle {\cal O}^{h_{c}} ({}^{1}P_1^{[1]}) } 
                \rangle \over 8} ) \cr
                &+& \biggl\lbrack  
                  \hat F_{{}^{3}S_1^{[8]}} 
                 + \hat F_{{}^{1}P_1^{[8]}} 
                + \hat F_{{}^{1}S_0^{[8]}} 
		+ \hat F_{{}^{3}P_J^{[8]}}  \biggr\rbrack 
		\times \langle {\cal O}^{h_c}  \rangle ,
\label{modified1}
\end{eqnarray}
where
\begin{equation}
     \langle {\cal O}^{h_c}  \rangle =
                \times \biggl\lbrack { M^2
                 \langle {\cal O} ({}^{3}S_1^{[8]}) \rangle} 
                + {M^2 \langle {\cal O} ({}^{1}S_0^{[8]}) \rangle} 
                + {\langle {\cal O} ({}^{3}P_J^{[8]}) \rangle}
                    \biggr\rbrack. 
\end{equation}

In contrast to the NRQCD, we needed to fix three non-
perturbative parameters to get the $h_c$ cross-section, but 
in modified NRQCD,
it is the sum of these parameters: so we have a single parameter
to fit. Similar formulation is being made for $h_b$ production.

Using heavy-quark symmetry relations, the non-perturbative parameters for 
$h_c$ and $h_b$ can be determined from $\chi_c$ and $\chi_b$
respectively. We have used the following heavy-quark symmetry relations: 
\bea
{{\langle {\cal O}^{h_{c}} ({}^{1}P_1^{[1]})\rangle }}&=&
3 \times {{\langle {\cal O}^{\chi_{c0}} ({}^{3}P_J^{[1]})\rangle }}= 
0.321~GeV^5~\cite{Butenschoen:2012qr}, 
\\
{{\langle {\cal O}^{h_{c}} ({}^{1}S_0^{[8]})\rangle }}&=&
3 \times {{\langle {\cal O}^{\chi_{c0}} ({}^{3}S_1^{[8]})\rangle }}= 
0.0066~GeV^3~\cite{Butenschoen:2012qr},
\\
{{\langle {\cal O}^{h_{b}} ({}^{1}P_1^{[1]})\rangle }}&=&
3 \times {{\langle {\cal O}^{\chi_{b0}} ({}^{3}P_J^{[1]})\rangle }}= 
7.2~GeV^5~\cite{Braaten:2000cm}, 
\\
{{\langle {\cal O}^{h_{b}} ({}^{1}S_0^{[8]})\rangle }}&=&
3 \times {{\langle {\cal O}^{\chi_{b0}} ({}^{3}S_1^{[8]})\rangle }}= 
0.045~GeV^3~\cite{Braaten:2000cm},
\\
\left< {\cal O}^{h_c} \right>&=& 
\left< {\cal O}^{\chi_c} \right>,\\
\left< {\cal O}^{h_b} \right>&=& 
\left< {\cal O}^{\chi_b} \right>, 
\eea

where $\left< {\cal O}^{\chi_c} \right>$  
and $\left< {\cal O}^{\chi_b} \right>$
are the fitted parameters for $\chi_c$ and $\chi_b$ respectively. 
We have taken $\left< {\cal O}^{\chi_c} \right> = - 0.0107~ \rm{GeV}^5$~\cite{bms2} 
and $\left< {\cal O}^{\chi_b} \right> = - 0.3~ \rm{GeV}^5$~\cite{Biswal:2023xnk},  
which were obtained earlier using modified NRQCD.
The matrix elements for the subprocesses are listed in 
Refs.~\cite{Cho:1995ce, Gastmans:1987be}.

\begin{figure}[ht!]
\begin{center}
\includegraphics[width=16cm]{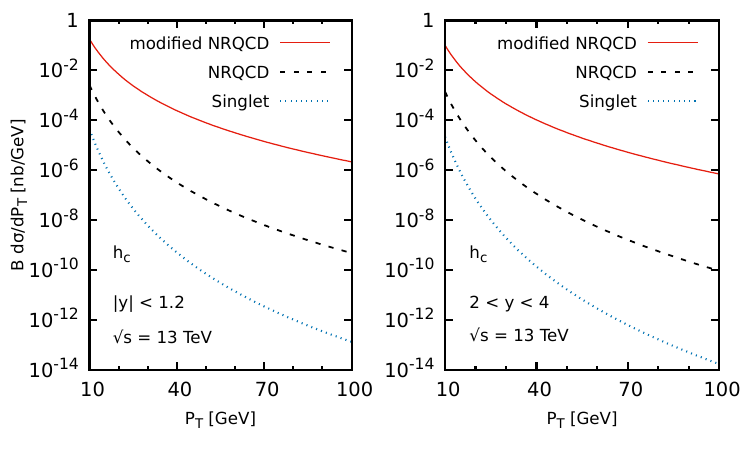}
\caption{Differential cross-section for $h_c$ production at the 13 TeV LHC.}
        \label{fig:fig1}
\end{center}
\end{figure}
\begin{figure}[ht!]
\begin{center}
\includegraphics[width=16cm]{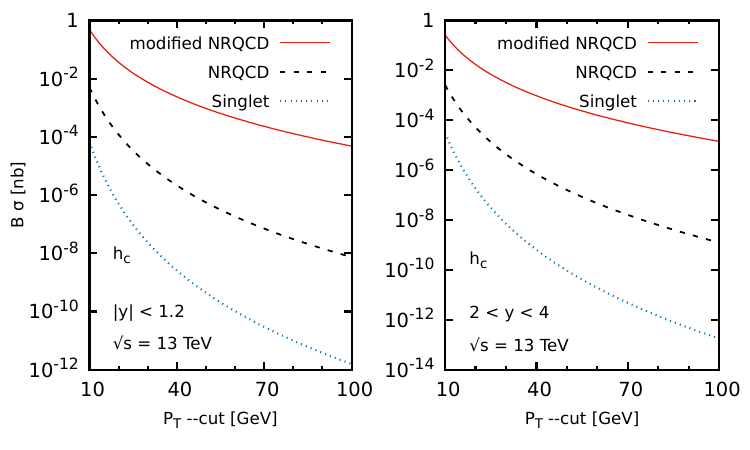}
\caption{Integrated cross-section for $h_c$ production at the 13 TeV LHC.}
        \label{fig:fig2}
\end{center}
\end{figure}

\begin{figure}[ht!]
\begin{center}
\includegraphics[width=16cm]{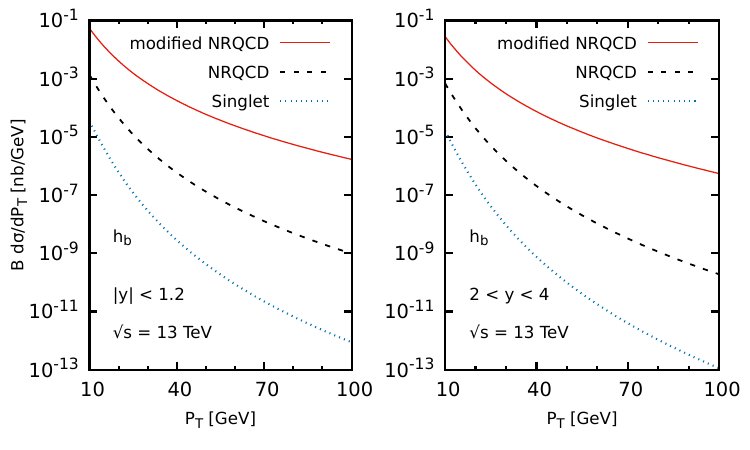}
\caption{Differential cross-section for $h_b$ production at the 13 TeV LHC.}
        \label{fig:fig3}
\end{center}
\end{figure}
\begin{figure}[ht!]
\begin{center}
\includegraphics[width=16cm]{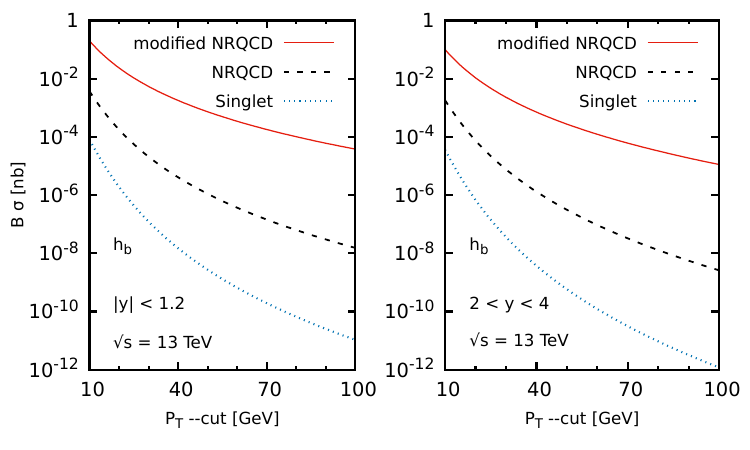}
\caption{Integrated cross-section for $h_b$ production at the 13 TeV LHC.}
        \label{fig:fig4}
\end{center}
\end{figure}
Figs.~1 and 2 represent the $h_c$ production 
differential cross-sections as a function of $p_T$ and 
integrated cross-sections for different $p_T$-cuts in 
both modified NRQCD and NRQCD. 
Similarly, Figs. 3 and 4 are for $h_b$ production.
Here we study $h_c$ in its decay into $\eta_c$ and $\gamma$ 
with a branching fraction of 60$\%$ and $\eta_c$ decays 
into $p\overline{p}$ state with a branching fraction of $1.33 \times 10^{-3}$~\cite{pdg22}. 
For $h_b$ production, we consider its decay 
into $\eta_b$ and $\gamma$ taking the branching fraction of 52$\%$ and 
a $1.33 \times 10^{-3}$ $p\overline{p}$ decay branching fraction for 
$\eta_b${\footnote {we have taken the branching ratio of 
$\eta_c \to p \bar p$ as the branching ratio of $\eta_b \to p \bar p$ 
is not available}.
Singlet prediction is also shown in all the figures. 
As the modified NRQCD prediction is very different from
both NRQCD and singlet prediction, 
a cross-section measurement of the ${}^1P_1$ production at the LHC experiments
can provide a crucial test of these interesting set of predictions.
\begin{table}[htbp]
\centering
\begin{tabular}{ |m{3cm}|m{3cm}|m{3cm}|m{3cm}|m{cm}|m{3cm}| }
 \hline
 \multicolumn{1}{|c|}{ } &\multicolumn{ 1}{c|}{ } 
&\multicolumn{4}{c|}{}
 \\[2mm]
\multicolumn{1}{|c|}{ } &\multicolumn{ 1}{c|}{ } 
&\multicolumn{4}{c|}{$\sim$ Expected number of events}
 \\[2mm]
 \cline{3-6}
  \multicolumn{1}{|c|}{ } &\multicolumn{ 1}{c|}{ } 
&\multicolumn{2}{c|}{}
&\multicolumn{2}{c|}{}
\\[2mm]
 \multicolumn{1}{|c|}{ } &\multicolumn{ 1}{c|}{ } 
&\multicolumn{2}{c|}{$h_c$ production}
&\multicolumn{2}{c|}{$h_b$ production}
\\[2mm]
\cline{3-6}
\multicolumn{1}{|c|}{Rapidity range} 
&\multicolumn{ 1}{c|}{$p_{T}$ range (GeV)} 
&\multicolumn{1}{c|}{}&\multicolumn{1}{c|}{}
&\multicolumn{1}{c|}{}&\multicolumn{1}{c|}{}
 \\[2mm]
\multicolumn{1}{|c|}{} 
&\multicolumn{ 1}{c|}{} 
&\multicolumn{1}{c|}{NRQCD}&\multicolumn{1}{c|}{Modified}
&\multicolumn{1}{c|}{NRQCD}&\multicolumn{1}{c|}{Modified}
 \\[2mm]
 \multicolumn{1}{|c|}{ } &\multicolumn{ 1}{c|}{ } 
&\multicolumn{1}{c|}{prediction}
&\multicolumn{1}{c|}{NRQCD}
&\multicolumn{1}{c|}{prediction}
&\multicolumn{1}{c|}{NRQCD}
 \\[2mm]
  \multicolumn{1}{|c|}{ } &\multicolumn{ 1}{c|}{ } 
&\multicolumn{1}{c|}{}
&\multicolumn{1}{c|}{prediction}
&\multicolumn{1}{c|}{}
&\multicolumn{1}{c|}{prediction}
 \\[2mm]
 \hline \hline
\multicolumn{1}{|c|}{2 $<$ y $<$ 4} 
&\multicolumn{ 1}{c|}{5 $<$ $p_{T}$ $<$ 15} 
&\multicolumn{1}{c|}{$1.4 \times 10^5$}
&\multicolumn{1}{c|}{$6.6 \times 10^6$}
&\multicolumn{1}{c|}{$2.9 \times 10^4$}&\multicolumn{1}{c|}{$9.1 \times 10^5$}
 \\[2mm]
 \hline
\multicolumn{1}{|c|}{-1.2 $<$ y $<$ 1.2} 
&\multicolumn{ 1}{c|}{10 $<$ $p_{T}$ $<$ 100} 
&\multicolumn{1}{c|}{$1.0 \times 10^4$}&\multicolumn{1}{c|}{$9.8 \times 10^5$}
&\multicolumn{1}{c|}{$7.5 \times 10^3$}&\multicolumn{1}{c|}{$4.0 \times 10^5$}
 \\[2mm]
 \hline 
\end{tabular}
\caption{Number of ${}^1P_1$ production events expected at the 
LHC running at $\sqrt{s}$ = 13 TeV.}
\label{table1}
\end{table}
To get a sense of the feasibility of measuring the 
$h_c$ and $h_b$ production at the LHC, 
we have also calculated the $p_T$-integrated cross-sections 
in two different rapidity ranges.
These numbers are presented in table \ref{table1} 
for an integrated luminosity of $2\ {\rm fb}^{-1}$ and 
suggest that one should expect a sizeable
number of $h_c$ and $h_b$ events at the LHC experiments.

The LHCb experiment has recently published some 
new results on integrated cross-section
for $h_c$ production \cite{LHCb:2024ydi}
at $\sqrt{s}$ = 13 TeV in the kinematic range 
5.0 $<$ $p_T$ $<$ 20.0 GeV and 2.0 $<$ $y$ $<$ 4.0. 
Therefore, we have compared the 
LHCb results for $h_c(1P)$ prompt production
cross-section with theoretical prediction
using both NRQCD and modified NRQCD.
We have obtained an integrated cross-section
for $h_c$ production using NRQCD is 
0.007 nb, while the modified NRQCD approach gives 
the value of 0.371 nb.
Moreover, we have observed that the predictions
of $h_c$ production
for modified NRQCD is in agreement
with the measured LHCb data.

In conclusion, 
we have studied the $^{1}P_1$ quarkonia production in 
both modified NRQCD and NRQCD, using the heavy-quark symmetry of NRQCD. 
As a model discriminating observable, 
we suggest the measurements of the integrated cross-section and 
the $p_T$ distribution of the $^{1}P_1$ quarkonia production at the LHC. 
We show that there are huge differences in the 
integrated cross-section and the $p_T$ distribution 
of $^{1}P_1$ quarkonia in the two models. 
These resonances should be measured in order to distinguish 
between modified NRQCD and NRQCD and to provide additional 
insight into the dynamics of quarkonium production.
Furthermore, while comparing with the recent data
from the LHCb experiment for 
the integrated cross-section
of $h_c$ production with the theoretical 
predictions using NRQCD and modified NRQCD, 
we find that modified NRQCD gives an agreement
with data from the LHCb experiment. 



\section*{Acknowledgments}
One of us (K.S.) gratefully acknowledges a research grant 
(No. 122500) from the
Azim Premji University. In this paper, as in everything else he writes, 
the results, opinions and views expressed are K.S.'s own and 
are not that of the Azim Premji University.  
S. S. M. acknowledges support from the 
Science and Technology department, 
Government of Odisha under DST letter No 2666/ST.


\end{document}